\documentstyle[prd,aps]{revtex}

\input epsf
\epsfverbosetrue

\mathchardef\ddash="705C

\tighten
\draft
\begin{document} 
\bibliographystyle{prsty}

\title{
Radiatively Induced Neutrino Masses and Oscillations\\
in an $SU(3)_L \times U(1)_N$ Gauge Model
\footnote{to be published in Phys. Rev. D (2001).}
}

\author{
Teruyuki Kitabayashi$^a$
\footnote{E-mail:teruyuki@post.kek.jp}
and Masaki Yasu\`{e}
\footnote{E-mail:yasue@keyaki.cc.u-tokai.ac.jp}
}

\address{\vspace{5mm}$^a$
{\sl Accelerator Engineering Center} \\
{\sl Mitsubishi Electric System \& Service Engineering Co.Ltd.} \\
{\sl 2-8-8 Umezono, Tsukuba, Ibaraki 305-0045, Japan}
}
\address{\vspace{2mm}$^b$
{\sl Department of Natural Science\\School of Marine
Science and Technology, Tokai University}\\
{\sl 3-20-1 Orido, Shimizu, Shizuoka 424-8610, Japan\\and\\}
{\sl Department of Physics, Tokai University} \\
{\sl 1117 KitaKaname, Hiratsuka, Kanagawa 259-1292, Japan}}
\date{TOKAI-HEP/TH-0004, October, 2000}
\maketitle

\begin{abstract}
We have constructed an $SU(3)_L \times U(1)_N$ gauge model utilizing an $U(1)_{L^\prime}$ symmetry, where $L^\prime$ = $L_e-L_\mu-L_\tau$, which accommodates tiny neutrino masses generated by $L^\prime$-conserving one-loop and $L^\prime$-breaking two-loop radiative mechanisms.  The generic smallness of two-loop radiative effects compared with one-loop radiative effects describes the observed hierarchy of $\Delta m_{atm}^2$ $\gg$ $\Delta m_\odot^2$.  A key ingredient for radiative mechanisms is a charged scalar ($h^+$) that couples to charged lepton-neutrino pairs and $h^+$ together with the standard Higgs scalar ($\phi$) can be unified into a Higgs triplet as ($\phi^0$, $\phi^-$, $h^+$)$^T$. This assignment in turn requires lepton triplets ($\psi_L^i$) with heavy charged leptons ($\kappa_L^{+i}$) as the third member: $\psi_L^i=(\nu^i_L,\ell^i_L,\kappa^{+i}_L)^T$, where $i$ ($=1,2,3$) denotes three families. It is found that our model is relevant to yield quasi-vacuum oscillations for solar neutrinos.
\end{abstract}
\vspace{2mm}
\section{Introduction}

The Super-Kamiokande experiments on atmospheric neutrino oscillations provide evidence for neutrino masses and mixings \cite{Kamiokande,RecentSK}, which require new interactions beyond the standard model \cite{MassiveNeutrino}. Solar neutrinos are also observed to be oscillating \cite{SolarNeutrino}.  These neutrino oscillations are governed by two mass scales, which are characterized by $\Delta m_{atm}^2={\cal O}$($10^{-3}$) eV$^2$ and $\Delta m_\odot^2 \le {\cal O}$($10^{-5}$) eV$^2$, respectively, suggested by atmospheric and solar neutrino oscillation data.  Furthermore, the recent analysis on atmospheric neutrino oscillations done by the K2K collaboration \cite{RecentK2K} has shown that $\Delta m_{atm}^2 \approx 3 \times 10^{-3}$ eV$^2$ well reproduces their data.  The atmospheric neutrino data, thus, imply 5.5$\times 10^{-2}$ eV as neutrino masses.  To generate such tiny neutrino masses, two main theoretical mechanisms have been proposed: one is the seesaw mechanism \cite{Seesaw} and the other is the radiative mechanism \cite{1-loop,2-loop,Radiative}. Recently, there have been various studies on implementation of radiative mechanisms in extended electroweak theories \cite{RecentRadiative}. In the radiative mechanism proposed by Zee \cite{1-loop}, a new singly charged $SU(2)_L$-singlet Higgs scalar, $h^+$, was introduced into the standard model and neutrino masses were generated as one-loop radiative corrections via the $h^+$-coupling to $\ell_L\nu_L$. After this work, Zee and Babu studied two-loop radiative mechanism \cite{2-loop}. One more doubly charged $SU(2)_L$-singlet Higgs scalar $k^{++}$ was added to the standard model and tiny neutrino masses arose from two-loop radiative effects initiated by $\ell_R\ell_R k^{++}$.

To account for the hierarchy of $\Delta m_{atm}^2$ $\gg$ $\Delta m_\odot^2$ suggested by the experimental data, it has been pointed out that bimaximal mixing scheme for neutrinos \cite{Mixing,NearlyBiMaximal} can $\ddash$algebraically" describe the hierarchy. One of the underlying physics behind bimaximal mixing is the presence of a new symmetry based on a lepton number of $L_e - L_\mu - L_\tau$ ($\equiv$ $L^\prime$) \cite{Lprime}.
\footnote{For earlier attempts of using such modified lepton numbers, see, for example, Ref.\cite{EarlierLprime}.}
A finite and very small $\Delta m_\odot^2$ is induce by a tiny breaking of the $L^\prime$-conservation. When this symmetry is combined with radiative mechanisms, the hierarchy can be $\ddash$dynamically" ascribed to the generic smallness of two-loop radiative effects compared with one-loop radiative effects \cite{1loop2loop,1loop2loopNew}.  Namely, one-loop radiative effects ensure the generation of $\Delta m_{atm}^2$ while the finite but small amount of $\Delta m_\odot^2$ is induced by two-loop radiative effects, which should involve $L^\prime$-breaking interactions \cite{1loop2loopNew}.
\footnote{For two-loop radiative mechanism based on $SU(2)_L\times U(1)_Y$ utilizing the $U(1)_{L^\prime}$ symmetry, see Ref.\cite{SU2U1}.}
In general, one-loop effects are much greater than two-loop effects, so the observed mass hierarchy between $\Delta m_{atm}^2$ and $\Delta m_\odot^2$ is understood to be based on this difference.

In this paper, along this line of thought on the realization of the hierarchy between $\Delta m_{atm}^2$ and $\Delta m_\odot^2$, we focus on studying neutrino physics characterized by radiative mechanisms.  In the one-loop radiative mechanism, interactions of $h^+$ provide breakdown of the lepton number conservation, which is an essential source for generating Majorana neutrino masses.  Now, we can ask what are the candidates of the $h^+$-scalar? There seems to be  many possibilities to answer this question. For example, $h^+$ can be identified with 1) the scalar (anti-)leptons ($\tilde{\ell^c}$) in SUSY theories, $h^+ = \tilde{\ell^c}$ \cite{SUSY} or 2) the third member of enlarged $SU(2)_L$ Higgs scalar in $SU(3)_L$ models \cite{SU3U1,SU3U1Topics}, $(\phi^0,\phi^-) \to (\phi^0,\phi^-, h^+)$ \cite{Okam99}. Here, we will examine phenomena of neutrino oscillations based on the second idea. 
One of the authors (M.Y.) has discussed theoretical framework for the radiative neutrino mass generation based on one-loop effects in various $SU(3)_L \times U(1)_N$ models \cite{Okam99}. We further seek possible mechanisms for the radiative neutrino mass generation based on both one-loop and two-loop effects in the $SU(3)_L \times U(1)_N$ framework. Among various versions of models,
\footnote{For a model with heavy neutral leptons corresponding to the choice of ($\phi^+$, $\phi^0$, $h^+$), see Ref.\cite{Kita00}.}
the present discussions concern the implementation of one- and two-loop radiative mechanisms in a model with heavy charged leptons ($\kappa^{+i}$),
\footnote{This charge assignment is identical to that of $\ell^c$.  However, one-loop radiative mechanism cannot work in this case.  See the details in Ref. \cite{Okam99}.}
 namely, with ($\nu^i_L$, $\ell^i_L$, $\kappa^{+i}_L$) for $i$=1,2,3, which is consistent with the choice of ($\phi^0$, $\phi^-$, $h^+$).

\section{Model}

The $SU(3)_L$ $\times$ $U(1)_N$ model with heavy charged leptons in the third member of lepton triplets is specified by the $U(1)_N$-charge.  Let $N/2$ be the $U(1)_N$ quantum number, then the hypercharge, $Y$, is given by $Y=-\sqrt{3}\lambda^8+N$ and the electric charge $Q_{em}$ is given by $Q_{em}=(\lambda^3+Y)/2$, where $\lambda^a$ is the $SU(3)$ generator with Tr$(\lambda^a \lambda^b)=2\delta^{ab}$ $(a,b=1...8)$.  The pure $SU(3)_L$-anomaly is cancelled in a vectorial manner.  The anomalies from triplets of the three families of leptons and of the three colors of the first family of quarks are cancelled by those from anti-triplets of three colors of the second and third families of quarks \cite{SU3U1}.   Other anomalies including $U(1)_N$ are also cancelled.  

Summarized as follows is the particle content in our model, where the quantum numbers for the $SU(3)_L$ and $U(1)_N$ symmetry are placed in the parentheses:
\begin{equation}
\psi^{i=1,2,3}_L = 
    \left( 
        \begin{array}{c}
        \nu^i_L     \\
        \ell^i_L    \\
        \kappa^{i}_L\\
        \end{array}  
    \right)    : \left( \textbf{3}, 0 \right), \quad
\ell^  {1,2,3}_R : \left( \textbf{1},-1 \right), \quad  
\kappa^{1,2,3}_R : \left( \textbf{1}, 0 \right),
\label{Eq:Leptons}
\end{equation}
for leptons, where we have denoted $\kappa^{+i}$ by $\kappa^i$,
\begin{eqnarray}
Q^1_L = 
    \left(
        \begin{array}{c}
        u^1_L \\
        d^1_L \\
        J^1_L \\
        \end{array}
    \right) : \left( \textbf{3}, \frac{2}{3} \right), \quad
Q^{i=2,3}_L = 
    \left(
        \begin{array}{c}
         d^i_L \\
        -u^i_L \\
         J^i_L \\
        \end{array}
    \right) : \left( \textbf{3}^\ast, -\frac{1}{3} \right),
\nonumber \\
u^{1,2,3}_R : \left( \textbf{1}, \frac{2}{3} \right),\quad 
d^{1,2,3}_R : \left( \textbf{1},-\frac{1}{3} \right),\quad 
J^1      _R : \left( \textbf{1}, \frac{5}{3} \right),\quad 
J^{2,3  }_R : \left( \textbf{1},-\frac{4}{3} \right),
\label{Eq:Quarks}
\end{eqnarray}
for quarks, and
\begin{eqnarray}
&& \eta =  
       \left(
           \begin{array}{c}
           \eta^0 \\
           \eta^- \\
           \eta^+ \\
           \end{array} 
       \right) : \left( \textbf{3}, 0 \right), \quad
    \rho = 
       \left(
           \begin{array}{c}
           \rho^+    \\
           \rho^0    \\
           \rho^{++} \\
           \end{array} 
       \right) : \left( \textbf{3}, 1 \right),
\nonumber \\
&&  \chi = 
       \left(
           \begin{array}{c}
           \chi^-    \\
           \chi^{--} \\
           \chi^0    \\
           \end{array} 
       \right) : \left( \textbf{3}, -1 \right),
\quad
    k^{++} : (\textbf{1},2),
\label{Eq:HiggsTriplet}
\end{eqnarray}
for Higgs scalars, where $\eta^+$ is Zee's $h^+$. Denoted by $k^{++}$ is an additional doubly charged $SU(3)_L$-singlet Higgs scalar responsible for two-loop radiative mechanism. The Higgs scalars have the following vacuum expectation values (VEV's):
\begin{equation}
<0|\eta|0> = 
    \left( \begin{array}{c}    
           v_\eta \\
           0      \\
           0      \\
           \end{array}
    \right), \quad 
<0|\rho|0> = 
    \left( \begin{array}{c}    
           0      \\
           v_\rho \\
           0      \\
           \end{array}
    \right), \quad 
<0|\chi|0> = 
    \left( \begin{array}{c}    
           0      \\
           0      \\
           v_\chi \\
           \end{array}
    \right),
\label{Eq:VEV}
\end{equation}
and quarks and leptons will acquire masses via these VEV's, where the orthogonal choice of these VEV's is ensured by the interaction of the $\eta\rho\chi$-type to be introduced in Eq.(\ref{Eq:HiggsPotential}). 

The conventional two-loop radiative mechanism in $SU(2)_L$ $\times$ $U(1)_Y$ is onset by $h^+h^+k^{++\dagger}$ together with $\ell^i_L\nu^j_Lh^+$.   Since both $\nu^i_L$ and $\ell^i_L$ are contained in a lepton triplet, $\psi^i_L$, $\ell^i_L\nu^j_Lh^+$ is replaced by $\psi^i_L\psi^j_L\eta$.  And $h^+h^+k^{++\dagger}$ can be replaced by $\eta^+\eta^+k^{++\dagger}$ in our notations.  By a possible $k^{++}$-$\rho^{++}$ mixing induced after $SU(3)_L$ $\times$ $U(1)_N$ is spontaneously broken to $U(1)_{em}$, the $\eta^+\eta^+k^{++\dagger}$ can be further rewritten as $\eta^+\eta^+\rho^{++\dagger}$ and is finally converted into the $SU(3)_L$ $ \times$ $U(1)_N$ - invariant form of $(\chi^\dagger \eta)(\rho^\dagger \eta)$, where 
$\langle 0 | \chi^0 | 0 \rangle$ $\neq$ 0 recovers $\eta^+\eta^+\rho^{++\dagger}$.  Similar argument also reveals that $\langle 0 | \rho^0 | 0 \rangle$ $\neq$ 0 recovers $\eta^-\eta^-\chi^{--\dagger}$, which turns out to be $\eta^-\eta^-k^{++}$ via a possible $k^{++}$-$\chi^{--\dagger}$ mixing, where $\kappa^i\nu^j_L\eta^-$ is used instead of $\ell^i_L\nu^j_L\eta^+$.  It is further obvious that an $SU(3)_L\times U(1)_N$-invariant $\rho^\dagger \chi k^{++}$ yields the necessary mixings with $k^{++}$.  Therefore, we employ  $(\chi^\dagger \eta)(\rho^\dagger \eta)$ and $\rho^\dagger \chi k^{++}$ as our new Higgs interactions.

We impose the $L^\prime$-conservation on our interactions to reproduce the observed atmospheric neutrino oscillations. This new lepton number, $L^\prime$, is assigned to the participating particles as shown in Table 1 together with the assignment of the lepton number, $L$.  There should be tiny breakdown of the $L$- and $L^\prime$-conservations, which, respectively, induce Majorana neutrino masses and the observed solar neutrino oscillations.  In the present case, the interactions of $\psi^i_L\psi^j_L\eta$

\footnote{Since there are quark mass terms such as ${\overline {Q^1_L}}\eta u^1_R$, the $L$-charge of $\eta$ should be set to zero.}
and $\rho^\dagger \chi k^{++}$, respectively, provide tiny breakings of the $L$- and $L^\prime$-conservations. Other interactions are assumed to satisfy the $L$- and $L^\prime$-conservations. 

The Higgs interactions are given by self-Hermitian terms: $\eta_\alpha \eta_\beta^\dagger, \rho_\alpha \rho_\beta^\dagger, \chi_\alpha \chi_\beta^\dagger$ and $k^{++}_\alpha  k_\beta^{++\dagger}$ and by two types of non-self-Hermitian Higgs potentials, $L^\prime$-conserving potential ($V_0$) involving the $(\chi^\dagger \eta)(\rho^\dagger \eta)$-term and $L^\prime$-violating potential ($V_b$) for the $\rho^\dagger \chi k^{++}$-term: 
\begin{eqnarray}
V_0 &=& \lambda_0 \epsilon^{\alpha\beta\gamma} \eta_\alpha \rho_\beta \chi_\gamma
      + \lambda_1 (\chi^\dagger \eta)(\rho^\dagger \eta)
      + (h.c.),
\nonumber \\
V_b &=& \mu_b \rho^\dagger \chi k^{++} + (h.c.),
\label{Eq:HiggsPotential}
\end{eqnarray}
where $\lambda_{0,1}$ are the coupling constants and $\mu_b$ denotes the $L^\prime$-breaking mass scale. These three terms in Eq.(\ref{Eq:HiggsPotential}) exhaust all possible $SU(3)_L$ $\times$ $U(1)_N$-invariant non-self-Hermitian terms.  The $L^\prime$-conserving Higgs potential, $V_0$, is responsible for the bimaximal mixing and the $L^\prime$-violating potential, $V_b$, is responsible for tiny breaking of the bimaximal mixing structure. Experimental data suggest that neutrinos exhibit bimaximal mixing and this feature is ascribed to the existence of the $L^\prime$-conservations \cite{Lprime}.  If there is only the $L^\prime$-conserving Higgs potential, the eigenvalues of neutrino mass matrix are given by $0$ and $\pm m_\nu$ ($m_\nu$: neutrino mass) and, from these eigenvalues, we can describe only atmospheric neutrino oscillations. However, if the $L^\prime$-violating Higgs potential also exists, we can realize two-loop radiative mechanism and we successfully obtain both atmospheric and solar neutrino oscillations.

The Yukawa interactions, which involve the $\psi_L\psi_L\eta$-term, relevant for the neutrino mass generation are given by the following lagrangian:
\begin{eqnarray}
-{\mathcal{L}}_Y &=& 
     \frac{1}{2}\epsilon^{\alpha\beta\gamma}\sum_{i=2,3}f_{[1i]}
     \overline{\left(\psi_{\alpha L}^1 \right)^c} \psi_{\beta L}^i \eta_\gamma
   + \sum_{i=1,2,3} \overline{\psi_L^i}
     \left( f_\ell^i \rho \ell_R^i + f_\kappa^i \chi \kappa_R^i \right)
\nonumber \\
&& + f_k^{11} \overline{(\kappa_R^1)^c} k^{++\dagger} \kappa_R^1
   + (h.c.),
\label{Eq:Yukawa}
\end{eqnarray}
where $f$'s are Yukawa couplings with the relation $f_{[ij]}=-f_{[ji]}$ demanded by the Fermi statistics. Their explicit form is given by   
\begin{eqnarray}
&&\sum_{i=2,3}f_{[1i]} \Biggl[
      \left(\overline{\kappa_R^{c1}}\ell_L^i  -\overline{\ell_R^{c1}}  \kappa_L^i \right)\eta^0
    + \left(\overline{\nu_R^{c1}}   \kappa_L^i-\overline{\kappa_R^{c1}}\nu_L^i    \right)\eta^-
    + \left(\overline{\ell_R^{c1}}  \nu_L^i   -\overline{\nu_R^{c1}}   \kappa_L^i \right)\eta^+ 
    \Biggl]
\nonumber \\
&&  + \sum_{i=1,2,3} \Biggl[ f_\ell^i \left(
      \overline{\nu_L^i}\rho^+ + \overline{\ell_L^i}\rho^0+\overline{\kappa_L^i}\rho^{++}
      \right)\ell_R^i
    + f_\kappa^i \left( 
      \overline{\nu_L^i}\chi^- + \overline{\ell_L^i}\chi^{--}+\overline{\kappa_L^i}\chi^0
      \right) \kappa_R^i \Biggl]
\nonumber \\
&&  + f_k^{11}\overline{\kappa_L^{c1}} k^{++\dagger} \kappa_R^1
    + (h.c.).
\label{Eq:YukawaExplicit}
\end{eqnarray}
The possible interactions of $\Sigma_{i,j=2,3} f_k^{ij} \overline{(\ell_R^i)^c} k^{++} \ell_R^j$ are forbidden by the $L$-conservation.  

\section{Neutrino  masses and oscillations}

Now, we can discuss how radiative corrections induce neutrino masses in our model. The combined use of Yukawa interactions with the $L^\prime$-conserving Higgs potential, $V_0$, yields one-loop diagrams for Majorana neutrino mass terms as shown in Fig.1. These one-loop diagrams correspond to the following interactions
\begin{equation}
  \left( \eta^\dagger \psi_L^1     \right)
      \epsilon^{\alpha\beta\gamma} \rho_\alpha \chi_\beta \psi_{\gamma L}^{2,3}
+ \left( \eta^\dagger \psi_L^{2,3} \right)
      \epsilon^{\alpha\beta\gamma} \rho_\alpha \chi_\beta \psi_{\gamma L}^1.
\label{Eq:EffectiveOneLoop}
\end{equation}
We also obtain two-loop diagrams involving the $L^\prime$-violating Higgs potential, $V_b$, as shown in Fig.2, which correspond to an effective coupling of 
\begin{equation}
\epsilon^{\alpha\beta\gamma} \epsilon^{\delta\epsilon\zeta}
    \psi_{\alpha L}^i \rho_\beta \chi_\gamma \psi_{\delta L}^j \rho_\epsilon \chi_\zeta
\label{Eq:EffectiveTowLoop}
\end{equation}
with $i,j=2,3$. 

From one-loop diagrams, we calculate the following Majorana masses to be:
\begin{eqnarray}
m_{1i}^{(1)} &=& f_{[1i]}
    \lambda_1 
    \Biggl[ 
    \frac{
        m_{\ell i}^2 F \left(m_{\ell i}^2,m_{\eta +}^2,m_{\rho +}^2 \right)
      - m_{\ell 1}^2 F \left(m_{\ell 1}^2,m_{\eta +}^2,m_{\rho +}^2 \right)
    }
    {v_\rho^2}
\nonumber \\
&& +\frac{
        m_{\kappa i}^2 F \left(m_{\kappa i}^2,m_{\eta -}^2,m_{\chi -}^2 \right)
      - m_{\kappa 1}^2 F \left(m_{\kappa 1}^2,m_{\eta -}^2,m_{\chi -}^2 \right)
    }
    {v_\chi^2}
    \Biggl] v_\eta v_\rho v_\chi,
\label{Eq:MajoranaOneLoop}
\end{eqnarray}
where 
\begin{equation}
F(x,y,z)=\frac{1}{16\pi^2}
    \Biggl[\frac{x\log x}{(x-y)(x-z)}+\frac{y\log y}{(y-x)(y-z)}+\frac{z\log z}{(z-y)(z-x)}\Biggl],
\label{Eq:Fxyz}
\end{equation}
and, from two-loop diagrams, we find
\begin{eqnarray}
m_{ij}^{(2)} &=& -2\lambda_1f_{[1i]}f_k^{11}f_{[1j]}  \mu_bm_{\kappa 1}^2 v^2_\rho I_\kappa,
\label{Eq:MajoranaTwoLoop}
\end{eqnarray}
where $I_\kappa$ denotes the two-loop integral via heavy charged leptons in Fig.2 (a). From the Appendix, one observes that $I_\kappa$ is calculated to be: 
\begin{eqnarray}
&& I_\kappa = \frac{1}{m_\chi^2-m_k^2} \left[ J(m^2_\chi)-J(m^2_k) \right],
\nonumber \\
&& J(m^2)   = \frac{1}{m^2}G^2(m_\kappa^2,m_\eta^2,m^2),
\nonumber \\
&& G(x,y,z) = \frac{1}{16\pi^2} \frac{x \ln (x/z) - y \ln (y/z)}{x-y},
\label{Eq:I_J_G}
\end{eqnarray}
where the explicit form of $J$ is subject to the condition of $z$ $\gg$ $x$, $y$ in $G$.  This condition is 
fulfilled because $m_\chi^2$, $m_k^2$ $\gg$ $m^2_\eta$, $m^2_\kappa$ can be safely chosen in the present case. 

Now, we obtain the following neutrino mass matrix:
\begin{equation}
M_\nu =   
    \left(
    \begin{array}{ccc}
         0 & m_{12}^{(1)} & m_{13}^{(1)} \\
         m_{12}^{(1)} & m_{22}^{(2)} & m_{23}^{(2)} \\
         m_{13}^{(1)} & m_{23}^{(2)} & m_{33}^{(2)} \\
    \end{array}
    \right),
\label{Eq:MassMatrix}
\end{equation}
from which we find
\begin{eqnarray}
&& \Delta m_{atm}^2 = m_{12}^{(1)2}+m_{13}^{(1)2} \quad (=m_\nu^2), 
\nonumber \\
&& \Delta m_\odot^2 = 2 \left(  m_{22}^{(2)} \cos^2 \vartheta_{atm} 
                             + 2m_{23}^{(2)} \cos\vartheta_{atm} \sin\vartheta_{atm}
                             +  m_{33}^{(2)} \sin^2 \vartheta_{atm}
                        \right) m_\nu,
\label{Eq:DeltaM-CM}
\end{eqnarray}
where $\vartheta_{atm}$ is a mixing angle for $\nu_\mu$ and $\mu_\tau$ with $\cos\vartheta_{atm} = m_{12}^{(1)}/m_\nu$ and $\sin\vartheta_{atm} = m_{13}^{(1)}/m_\nu$.  This form of mass matrix shows that solar neutrino oscillations exhibit almost maximal mixing while atmospheric neutrino oscillations are characterized by the mixing angle of $\vartheta_{atm}$.   The bimaximal mixing is reproduced by requiring that $\sin 2\vartheta_{atm}$ $\approx$ $1$, namely, $m_{12}^{(1)} \approx m_{13}^{(1)}$. It is realized by $m_{\kappa 2} \sim m_{\kappa 3}$ or by $m_{\kappa 2}, m_{\kappa 3} \ll m_{\kappa 1}$ since the contributions from the charged-lepton-exchanges become more suppressed than those from the heavy-charged-lepton-exchanges. Here, we use ($m_{\kappa 1} \sim) m_{\kappa 2} \sim m_{\kappa 3}$ to reproduce bimaximal mixing structure but with $m_{\kappa 1}$ $\neq$ $m_{\kappa i}$ ($i$ = 2,3) to have non-vanishing contributions from the heavy-lepton exchanges.


In order to see whether our model gives the compatible description of neutrino oscillations with the observed data, we must specify various parameters in our model. We make the following assumptions on relevant free parameters to compute $\Delta m_{atm}^2$ and $\Delta m_\odot^2$ in Eq.(\ref{Eq:DeltaM-CM}):

\begin{enumerate}
\item 
	Since $v_{\eta,\rho}$ are related to masses of weak bosons proportional to $\sqrt{v^2_\eta+v^2_\rho}$, we require $\sqrt{v^2_\eta+v^2_\rho}$=$v_{weak}$, from which ($v_\eta$, $v_\rho$) = ($v_{weak}/20$, $v_{weak}$) are taken,
\footnote{Since the third family of quarks belongs to anti-triplet as in Eq.(\ref{Eq:Quarks}), the $t$-quark mass, $m_t$, is controlled by $v_\rho$, {\it i.e,} $m_t$ $\sim$ $v_\rho$ while $b$-quark mass, $m_b$, is controlled by $v_\eta$, {\it i.e.} $m_b$ $\sim$ $v_\eta /2$.}
where $v_{weak}$ = $( 2{\sqrt 2}G_F)^{-1/2}$ = 174 GeV, 
\item 
   Since $v_\chi$ is a source of masses for heavy charged leptons and also of masses for exotic quarks and exotic gauge bosons, we use $v_\chi$ $\gg$ $v_{weak}$, from which $v_\chi = 10v_{weak}$ is taken,
\item 
	The masses of the Higgs bosons, $\eta$ and $\rho$, are set to be $m_\eta = m_\rho = v_{weak}$,
\item 
	The masses of the Higgs bosons, $k^{++}$ and $\chi$, and of the heavy charged leptons, $\kappa^i$ ($i$ = 1,2,3), are assigned to be larger values as $m_k$ = $m_\chi$ = $v_\chi$ and $m_{\kappa 2, \kappa 3}$ = $ev_\chi$ supplemented by 10\% mass difference between $\kappa^1$ and $\kappa^{2,3}$, {\it i.e.} $m_{\kappa 1}$ = $0.9m_{\kappa 2, \kappa 3}$, where $e$ stands for the electromagnetic coupling,
\item 
	The $L$-violating couplings of $f_{[1i]}$ ($i$ = 2,3) are determined by $\Delta m^2_{atm}$, where $f_{[1i]}$ = $1.8 \times 10^{-7}$ is to be taken,
\item 
	The $L^\prime$-violating scale of $\mu_b$ is suppressed as $ev_\chi$,
\item 
	The $L$- and $L^\prime$-conserving couplings accompany no suppression factor and are set to be of order 1 as $f^k_{11}$ = $\lambda_1$ = 1.
\end{enumerate}
These values are tabulated in Table 2. 

One will find that this parameter set reproduces the correct magnitude of $\Delta m_{atm}^2$ and yields $\Delta m_\odot^2$ = ${\cal O}$(10$^{-9}$) eV$^2$ relevant to describe quasi-vacuum oscillations (QVO) for solar neutrinos,
\footnote{The VO solution with $\Delta m^2_{atm}$ = ${\cal O}$(10$^{-10}$) eV$^2$ can be supplied by using $v_\eta$ = $v_{weak}/10$ and $f_{[1i]}$ = 10$^{-7}$.  The recent report from the Super-Kamiokande collaboration has presented the statement that the VO solution seems to be disfavored at the 95\% confidence level \cite{RecentSK}. However, the VO solution may be still favored.  See, for example, Ref.\cite{VOComment}.}
which are similar to vacuum oscillations (VO) but with small matter corrections taken into account \cite{QVO0,QVO1,QVO2}. Let us first see the order of magnitude of $f_{[1i]}$ ($i$ = 2,3), which is about $10^{-7}$, determined so as to reproduce the observed $\Delta m_{atm}^2$.  The one-loop neutrino mass of $m^{(1)}_{1i}$ is roughly approximated to be:
\begin{eqnarray}
 m^{(1)}_{1i} \sim \frac{1}{2}f_{[1i]}\lambda_1\left( m^2_{\kappa i} - m^2_{\kappa 1}\right)\frac{1}{16\pi^2m^2_{\chi}}\frac{v_\eta v_\rho}{v_\chi},
\label{Eq:RoughEstimate1}
\end{eqnarray}
where $F$ $\sim$ $1/16\pi^2m^2_\chi$ has been used from the dimensional ground.  By inserting numerical values tabulated in Table 2, it is readily found that
\begin{eqnarray}
& m^{(1)}_{1i} \sim 10^5 f_{[1i]}~{\rm eV},
\label{Eq:RoughEstimate2}
\end{eqnarray}
which should reproduce $\sqrt{\Delta m^2_{atm}/2}$ $\sim$ $4 \times 10^{-2}$ eV.  Therefore, $f_{[1i]}$ $\sim$ $4 \times 10^{-7}$ is obtained. Next, through the similar argument on the estimation of the two-loop neutrino masses of $m^{(2)}_{ij}$ by using $I_\kappa$ $\sim$ $1/(16\pi^2)^2m^4_k$, one observes that
\begin{eqnarray}
& m^{(2)}_{ij} \sim 4\times 10^4 f_{[1i]} f_{[1j]}~{\rm eV}.
\label{Eq:RoughEstimate3}
\end{eqnarray}
This rough estimate yields $m^{(2)}_{ij}$ $\sim$ $6 \times 10^{-9}$ eV, which corresponds to $\Delta m^2_\odot$ $\sim$ $1.4\times 10^{-9}$ eV$^2$.  In fact, numerical analysis by using Gaussian integral method gives $f_{[ij]} = 1.8\times 10^{-7}$, which reproduces $\Delta m_{atm}^2$ = $2.6 \times 10^{-3}$ eV$^2$. The bimaximal structure is characterized by $\sin^2 2\vartheta_{atm}$ = 0.93, where the deviation from unity arises from the charged-lepton contributions to $m_\nu$. The solar neutrino oscillations are calculated to be $\Delta m^2_\odot$ = $8.1 \times 10^{-10}$ eV$^2$, which lies in the allowed region relevant to the QVO solution with almost maximal mixing. 

Two comments are in order.  The first one concerns the $L$- and $L^\prime$-charges of $k^{++}$.  There are three other choices of ($L$, $L^\prime$) of $k^{++}$, {i.e.} (2, $-$2), ($-$2, 2) and ($-$2, $-$2).  The similar conclusion can be obtained for the (2, $-$2) case.  The mass matrix of $M_\nu$ has the (1,1)-entry instead of the ($i$,$j$)-entry ($i$,$j$ = 2,3), which is induced by $\kappa^{2,3}_R\kappa^{2,3}_R k^{++\dagger}$.  Other cases yield too small $\Delta m^2_\odot$ $\sim$ ($m^2_\ell / m^2_\kappa$) $\times$ $10^{-9}$ eV$^2$.  The second one is about the inclusion of the charged-lepton-exchange effects in two-loop diagrams, which would arise without the $L$-conservation for the $k^{++}$-interactions.  If there were $\Sigma_{i,j=2,3} f_k^{ij} \overline{(\ell_R^i)^c} k^{++} \ell_R^j$,
\footnote{Of course, one has to worry about $k^{++}$-contributions to the well-established low-energy physics such as $\tau^-$ $\rightarrow$ $\mu^-\gamma$ and $\mu^-\mu^-\mu^+$.}
the (1,1) entry of $M_\nu$ of Eq.(\ref{Eq:MassMatrix}) would be $-2\lambda_1\Sigma_{i,j=2,3}$ $f_{[1i]}f_k^{ij}f_{[1j]}$ $\mu_bm_{\ell i}m_{\ell j}v^2_\chi$ $I(m^2_{\ell i}, m^2_{\ell j})$ (=$m_{11}^{(2)}$).  However, because of $m_{\ell i} v_\chi \ll m_{\kappa 1} v_\rho$ in the present case, $m^{(2)}_{11}$ turns out to be  more suppressed than $m^{(2)}_{ij}$.  In fact, the numerical computation gives $m^{(2)}_{11}$ =2.4 $\times$ $10^{-10}$ eV, which is much smaller than $m^{(2)}_{ij}$ = 4.1 $\times$ $10^{-9}$ eV ($i,j$ = 2,3).  So the inclusion of the two-loop effects via charged leptons does not alter our result.

\section{Summary}

We have constructed an $SU(3)_L \times U(1)_N$ gauge model, which provides the radiatively generated neutrino masses and the observed neutrino oscillations. The third member of each lepton triplet, $\psi_L^i$, is the heavy charged leptons, $\kappa_L^{+i}$, placed as $\psi_L^i$ = ($\nu^i_L$, $\ell^i_L$, $\kappa^{+i}_L$)$^T$. Similarly, the third member of the Higgs scalar, $\eta$, is Zee's $h^+$: $\eta$ = ($\phi^0$, $\phi^-$, $h^+$)$^T$ and the lepton-number-violating term is controlled by the interactions with this $\eta$ as 
$\psi^i_L\psi^j_L \eta$.  One-loop radiative neutrino masses are, then, induced by $\psi^i_L\psi^j_L\eta$ together with the Higgs interactions of $(\rho^\dagger\eta )(\chi^\dagger \eta )$.  Two-loop radiative mechanism calls for the doubly charged $k^{++}$, which has couplings to $(\kappa^1_R\kappa^1_R)^\dagger$. After the mixings of $k^{++}$ with $\rho^{++}$ and $\chi^{--\dagger}$ by $\rho^\dagger\chi k^{++}$, two-loop radiative neutrino masses are induced.

To account for the observed hierarchy between $\Delta m^2_{atm}$ $\gg$ $\Delta m^2_\odot$, we have utilized both the $L_e-L_\mu-L_\tau$ (=$L^\prime$) number conservation and more suppressed two-loop radiative corrections than one-loop radiative corrections.  The bimaximal mixing structure in $M_\nu$ is enhanced by the approximate degeneracy in masses of $\kappa^2$ and $\kappa^3$, {\it i.e.} $m_{\kappa 2}$ $\sim$ $m_{\kappa 3}$. 
The $L^\prime$-conservation ensures $\Delta m^2_{atm}$ $\gg$ $\Delta m^2_\odot$ (=0) with the maximal solar neutrino mixing.  Since $\Delta m^2_{atm}$ is given by one-loop radiative corrections and $\Delta m^2_\odot$ is given by two-loop radiative corrections, $\Delta m^2_{atm}$ $\gg$ $\Delta m^2_\odot$ ($\neq$ 0) is realized. The interactions involving $\eta^+$ respect the $U(1)_{L^\prime}$ symmetry while the interactions of $k^{++}$ contain a $L^\prime$-breaking term, which is supplied by $\rho^\dagger\chi k^{++}$.   The numerical estimate certainly provides $\Delta m^2_{atm}$ = $2.6 \times 10^{-3}$ eV$^2$ with $\sin^2 2\vartheta_{atm}$ = 0.93 and $\Delta m_\odot^2$ = $8.1 \times 10^{-10}$ eV$^2$. Our model is thus relevant to yield the QVO solution to the solar neutrino problem.

\begin{center}
{\bf ACKNOWLEDGMENTS}
\end{center}

The work of M.Y. is supported by the Grant-in-Aid for Scientific Research No 12047223 from the Ministry of Education, Science, Sports and Culture, Japan.


\begin{center}
{\bf APPENDIX}
\end{center}

In this Appendix, we describe the outline of obtaining Eq.(\ref{Eq:I_J_G}) from the two-loop integral 
corresponding to Fig.2 (b), which takes the form of 
\begin{equation}
I(m^2_c, m^2_d) = \int \frac{d^4k}{(2\pi)^4} \frac{d^4q}{(2\pi)^4}
        \frac{1}{k^2-m_c^2} \frac{1}{k^2-m_1^2} 
        \frac{1}{q^2-m_d^2} \frac{1}{q^2-m_3^2}    
        \frac{1}{\left[(k-q)^2-m_2^2 \right]} \frac{1}{\left[(k-q)^2-m_k^2 \right] }.
\label{Eq:Integral}
\end{equation}
The integral of $I_\kappa$ in Eq.(\ref{Eq:MajoranaTwoLoop}) is defined by $I_\kappa$ = $I(m^2_\kappa, m^2_\kappa)$.  By using the identity:
\begin{equation}
\frac{1}{(k-q)^2-m_2^2}\frac{1}{(k-q)^2-m_k^2} = \frac{1}{m_2^2-m_k^2}\left( \frac{1}{(k-q)^2-m_2^2}-\frac{1}{(k-q)^2-m_k^2} \right),
\label{Eq:identity}
\end{equation}
we can rewrite $I(m^2_c, m^2_d)$ to 
\begin{equation}
I=\frac{1}{m_2^2-m_k^2} \left( J(m^2_2)-J(m^2_k) \right),
\label{Eq:I}
\end{equation}
where 
\begin{equation}
J(m^2) \equiv \int\frac{d^4k}{(2\pi)^4}\frac{d^4q}{(2\pi)^4} \frac{1}{(q^2-m_d^2)(q^2-m_3^2)}
                  \frac{1}{(k^2-m_c^2)(k^2-m_1^2) \left[ (k-q)^2-m^2 \right]}.
\label{Eq:J0}
\end{equation}
By using the Feynman formula:
\begin{equation}
\frac{1}{abc} = \int_0^1 dx \int_0^1 2ydy \frac{1}{\left[ c+(b-c)y+(a-b)xy \right]^3},
\label{Eq:FeynmanIntegral}
\end{equation}
and by performing the 4-dimensional integrations over $k$ and $q$ with the aid of
\begin{equation}
\int \frac{d^4l}{(2\pi)^4}\frac{1}{(l-\Delta)^3}
    =-\frac{i}{32\pi^2}\frac{1}{\Delta},
\label{Eq:4dimIntegral}
\end{equation}
we reach
\begin{equation}
J(m^2) = -\left( \frac{1}{16\pi^2} \right)^2
    \Biggl[   \frac{m_d^2 \ln m_d^2}{m_d^2-m_3^2} I_1(m^2,m^2_d)
            + \frac{m_3^2 \ln m_3^2}{m_3^2-m_d^2} I_1(m^2,m^2_3)
            + I_2(m^2, m^2_3, m^2_d)
    \Biggl],
\label{Eq:Jm0}
\end{equation}
where
\begin{eqnarray}
&&I_1(m^2,m^2_A)    = \int_0^1dx\int_0^1ydy \frac{1}{-y(1-y) 
		\left[ m_A^2 - M(m^2) \right]},
\nonumber \\
&&I_2(m^2,m^2_A,m^2_B)=  \int_0^1dx\int_0^1ydy 
                    \frac{M(m^2) \ln M(m^2)}
                         {-y(1-y) \left[ m_A^2-M(m^2) \right]
										  \left[ m_B^2-M(m^2) \right]}, 
\nonumber \\
&&M(m^2)   = \frac{m^2-(m^2-m_1^2)y-(m_1^2-m_c^2)xy}
											{y(1-y)}.
\label{Eq:I1_I2_M}
\end{eqnarray}

The integral $I_1$ is calculated to be: 
\begin{equation}
I_1(m^2,m^2_A) = \frac{1}{m_c^2-m_1^2} Y(m^2, m_A^2),
\label{Eq:I1}
\end{equation}
after the integration over $x$, where 
\begin{equation}
Y(m^2, m_A^2) \equiv \int_0^1 dy 
						\ln \left| \frac{m_A^2y^2-(m_A^2+m^2-m_c^2)y+m^2}
                             			 {m_A^2y^2-(m_A^2+m^2-m_1^2)y+m^2}
 							   \right|.
\label{Eq:Y}
\end{equation}
This function, $Y(m^2, m_A^2)$, can be converted into
\begin{equation}
Y(m^2, m_A^2) = \int_0^1 dy \ln \left|\frac{(y-\alpha_+)(y-\alpha_-)}{(y-\beta_+)(y-\beta_-)}\right|,
\label{Eq:Y_no2}  
\end{equation}
where
\begin{equation}
\alpha_{\pm} = \frac{-a\pm\sqrt{a^2-4b}}{2}, \quad 
\beta_{\pm} = \frac{-c\pm\sqrt{c^2-4d}}{2}.
\label{Eq:AlphaBeta}  
\end{equation}
The parameters of $a$, $b$, $c$ and $d$ are defined by
\begin{eqnarray}
&&m_A^2y^2-(m_A^2+m^2-m_c^2)y+m^2 \equiv m^2_A \left( y^2+ay+b\right),
\nonumber \\
&&m_A^2y^2-(m_A^2+m^2-m_1^2)y+m^2 \equiv m^2_A \left( y^2+cy+d\right). 
\label{Eq:abcd}  
\end{eqnarray}
Therefore, $Y(m^2, m_A^2)$ is completely determined.  

The integral $I_2$ cannot be solved analytically.  However, if the approximation of 
$m^2$ $\gg$ $m^2_c$, $m^2_d$, $m^2_1$, $m^2_3$ is safely taken, the analytical expression of 
$I_2$ can be obtained. With the approximation, where $\ln M^2 \approx \ln m^2$  in $I_2$, we obtain
\begin{eqnarray}
I_2(m^2,m^2_A,m^2_B) &\approx& \int_0^1 dx\int_0^1 ydy 
        \frac{M(m^2) \ln m^2}
             {-y(1-y) \left[ m_A^2-M(m^2) \right] \left[ m_B^2-M(m^2) \right]},
\nonumber \\
 &=&    -\frac{\ln m^2}{m^2_A-m^2_B}\left(m^2_A I_1(m^2, m^2_A) - m^2_B I_1(m^2, m^2_B)\right)
\nonumber \\
 &=& 
	-\frac{\ln m^2}{(m_c^2-m_1^2)(m_A^2-m_B^2)} 
	\left( m_A^2 Y(m^2, m_A^2)  - m_B^2 Y(m^2, m_B^2) \right).
\label{Eq:I2}
\end{eqnarray}
Furthermore, $Y(m^2, m_A^2)$ can be approximated to be:
\begin{eqnarray}
& \alpha_+ \approx \frac{m^2}{m_A^2} \left( 1- \frac{m_c^2}{m^2} \right),
\quad
\alpha_-  \approx 1+ \frac{m_c^2}{m^2},
\label{Eq:alpha_beta} \\
& \beta_+ \approx \frac{m^2}{m_A^2} \left( 1- \frac{m_1^2}{m^2} \right),
\quad
\beta_-  \approx 1+ \frac{m_1^2}{m^2},
\label{Eq:gamma_delta}
\end{eqnarray}
from which $Y(m^2, m_A^2)$ is calculated to be:
\begin{equation}
Y(m^2, m_A^2) \approx -\frac{1}{m^2} 
	\left( m_c^2 \ln (m_c^2/m^2) - m_1^2 \ln (m_1^2/m^2) \right),
\label{Eq:Y_no3}
\end{equation}
which becomes independent of $m^2_A$.

Finally, from Eqs.(\ref{Eq:I}),(\ref{Eq:Jm0}),(\ref{Eq:I1}), (\ref{Eq:I2}) and (\ref{Eq:Y_no3}), we get
\begin{eqnarray}
&&I_\kappa = I(m^2_\kappa, m^2_\kappa) = \frac{J(m^2_\chi)-J(m^2_k)}{m_\chi^2-m_k^2},
\nonumber \\
&&J(m^2) = \frac{1}{m^2}G(m_\kappa^2,m_3^2,m^2)G(m_\kappa^2,m_1^2,m^2),
\nonumber \\
&&G(x,y,z) = \frac{1}{16\pi^2} \frac{x \ln (x/z) - y \ln (y/z)}{x-y},
\label{Eq:Ikappa_I_J_G}
\end{eqnarray}
which are the expressions in Eq.(\ref{Eq:I_J_G}) with the dominance of contributions from 
$m^2_\chi$ and $m^2_k$ in the integral. 


\noindent
\begin{center}
\textbf{Table Captions}
\end{center}
\begin{description}
\item{TABLE \ref{Table1}:} $L^\prime$ quantum number
\item{TABLE \ref{Table2}:} Model parameters, where masses are given in the unit of $v_{weak}$ = $( 2{\sqrt 2}G_F)^{-1/2}$ = 174 GeV and $e$ stands for the electromagnetic coupling.
\end{description}

\noindent
\begin{center}
\textbf{Figure Captions}
\end{center}

\begin{figure}
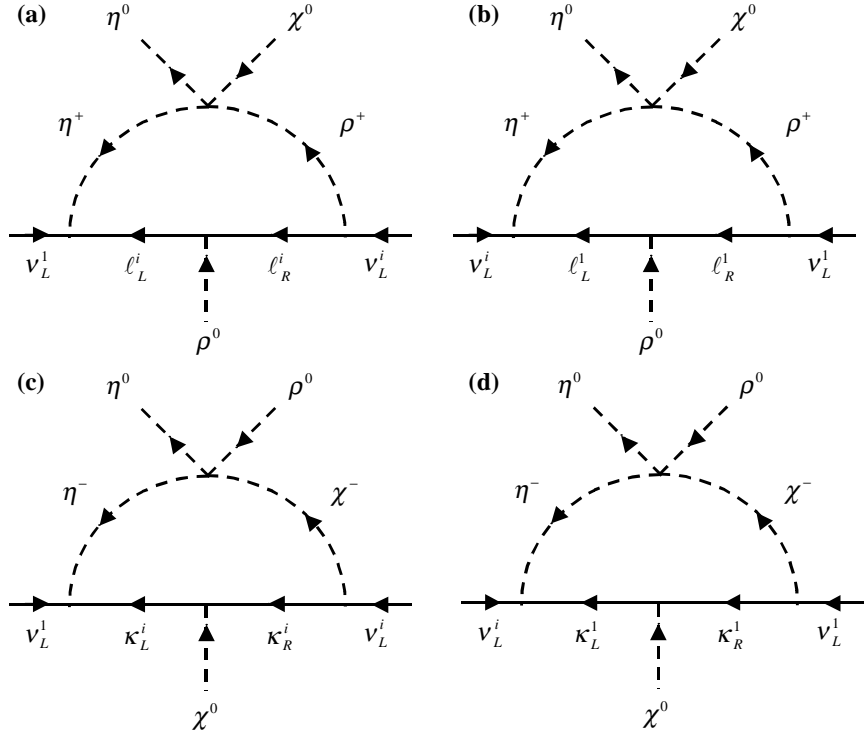

\caption{1-loop diagrams for $\nu^1-\nu^i~(i=2,3)$ via (a,b) charged leptons $\ell$ and via (c,d) heavy charged leptons $\kappa$.}
\label{Fig_1}
\end{figure}
\begin{figure}
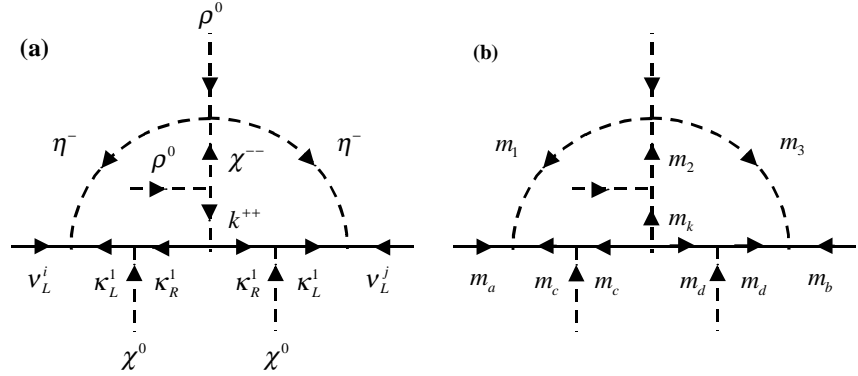

\caption{2-loop diagrams for (a) $\nu^i-\nu^j~(i,j=1,2)$ via heavy charged leptons $\kappa$ and for (b) notations used in loop calculations in Eq.(\ref{Eq:Integral}).}
\label{Fig_2}
\end{figure}

\begin{table}[ht]
    \caption{\label{Table1}$L^\prime$- and $L$-quantum numbers}
    \begin{center}
    \begin{tabular}{ccccc}
        Fields     & $\eta,\rho,\chi$                           & $\psi_L^1,\ell_R^1,\kappa_R^1$
        & $\psi_L^{2,3},\ell_R^{2,3},\kappa_R^{2,3}$ & $k^{++}$ \\
    \hline
        $L^\prime$ & 0 & 1 & $-$1 & 2 \\
    \hline
        $L$        & 0 & 1 & 1 & 2
    \end{tabular}
    \end{center}
\end{table}
\begin{table}[ht]
    \caption{\label{Table2}Model parameters, where masses are given in the unit of $v_{weak}$ = $( 2{\sqrt 2}G_F)^{-1/2}$ = 174 GeV and $e$ stands for the electromagnetic coupling.}
    \begin{center}
    \begin{tabular}{ccccccccccc}
     $v_\eta$ & $v_\rho$ & $v_\chi$ & $m_{k,\chi}$ & $m_{\eta,\rho}$ & $m_{\kappa 1}$ & $m_{\kappa 2.3}$ &$\mu_b$ &
     $\lambda_1$  & $f_k^{ij}$    & $f_{ij}$ \\
    \hline
       1/20 & 1 & 10   & 10 & 1 &   9$e$ &  10$e$  & 10$e$   &  1   & 1  & $1.8\times 10^{-7}$
    \end{tabular}
    \end{center}
\end{table}

\newpage
\epsfbox{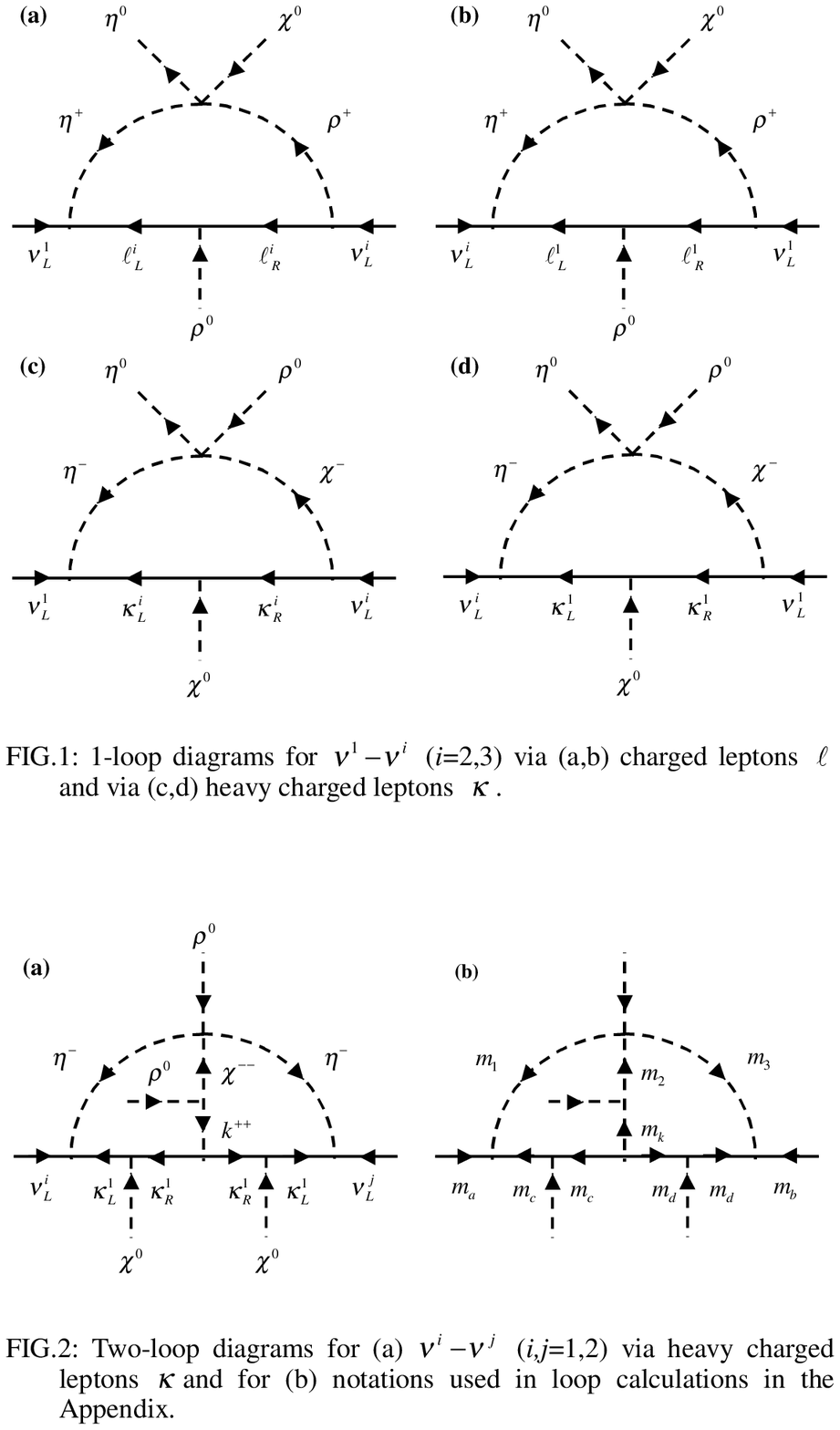}
\end{document}